\documentclass[prl,
showpacs,aps,nofootinbib,floatfix,amsmath,amssymb]{revtex4}
\usepackage{graphicx}
\usepackage{soul}
\usepackage{color}
\usepackage[usenames,dvipsnames]{xcolor}
\begin{document}

\makeatletter
\newbox\slashbox \setbox\slashbox=\hbox{$/$}
\newbox\Slashbox \setbox\Slashbox=\hbox{\large$/$}
\def\pFMslash#1{\setbox\@tempboxa=\hbox{$#1$}
  \@tempdima=0.5\wd\slashbox \advance\@tempdima 0.5\wd\@tempboxa
  \copy\slashbox \kern-\@tempdima \box\@tempboxa}
\def\pFMSlash#1{\setbox\@tempboxa=\hbox{$#1$}
  \@tempdima=0.5\wd\Slashbox \advance\@tempdima 0.5\wd\@tempboxa
  \copy\Slashbox \kern-\@tempdima \box\@tempboxa}
\def\FMslash{\protect\pFMslash}
\def\FMSlash{\protect\pFMSlash}
\def\miss#1{\ifmmode{/\mkern-11mu #1}\else{${/\mkern-11mu #1}$}\fi}

\newcommand{\psum}[1]{{\sum_{ #1}\!\!\!}'\,}
\makeatother

\title{The role of hidden symmetries and Kaluza-Klein mass generation in extra-dimensional gauge theories}
\author{H. Novales-S\'anchez and J. J. Toscano}
\affiliation{Facultad de Ciencias F\'isico Matem\'aticas, Benem\'erita Universidad Aut\'onoma de Puebla, Apartado Postal 1152 Puebla, Puebla, M\'exico}

\begin{abstract}
The transition from formulations with extra dimensions to Kaluza-Klein theories, aimed at extending the Standard Model, bears the ingredients of hidden symmetries and the Kaluza-Klein mechanism for mass generation. We explore these essential aspects in detail, and find that much can be said about them with no reference to the specific geometry of compact extra dimensions: the low-energy theory is determined, included dynamic variables and symmetries; mass terms arise; eigenfunctions that define Kaluza-Klein fields are fixed by the appropriate choice a Casimir invariant; there is a set of Kaluza-Klein pseudo-Goldstone bosons. Throughout our presentation, similarities and differences among spontaneous symmetry breaking, commonly present in conventional Standard Model extensions, and what happens in Kaluza-Klein theories are signaled and discussed.
\end{abstract}


\maketitle

It has been a long time since Kaluza and Klein first explored the possibility that the number of spacetime dimensions is greater than 4~\cite{ThK,OKlein}. After a silent long period, an increasing interest of the scientific community in extra dimensions started with its implementation in superstring theory~\cite{GrSch,JPolch,EWitt}. Shortly after, extra-dimensional phenomenology attracted great attention with the proposal of models of {\it large extra dimensions}~\cite{AADD,ADD}, {\it warped extra dimensions}~\cite{RaSu}, and {\it universal extra dimensions}~\cite{ACD}. Standard Model (SM) extensions in extra dimensions yield Kaluza-Klein (KK) theories, defined in terms of 4-dimensional dynamic variables and symmetries, which is achieved by a procedure that involves two main features: {\it hidden symmetries} and the mass-generating {\it Kaluza-Klein mechanism} (KKM). Subtle and complex scenarios of fundamental physics are elegantly described by hidden symmetries~\cite{LMNT}. Spontaneous symmetry breaking (SSB) merge with this powerful concept to become the very essence of the Englert-Higgs mechanism (EHM)~\cite{PHiggs,EngBr} and a cornerstone of the SM. The hiding of a symmetry is just a change of perspective, so it is implemented by canonical transformations, which, due to the Dirac algorithm~\cite{PDirac,GiTy,HenTei}, suggests that gauge symmetry should be preserved, at least, in the sense of a Lie group, though defined in a different spacetime and thus associated to different connections. Moreover, when passing to the 4-dimensional viewpoint, in which extra-dimensional spacetime has been hidden by some compactification scheme, some connections manifest as matter fields; with no protecting symmetry they can become massive. The assumption of compactness of extra dimensions generates masses, independently of the specific geometry of extra dimensions. A conspicuous difference among the EHM and the KKM is that mass generation in the former requires the introduction of extra degrees of freedom, which are independent of gauge symmetry, whereas the masses produced by the KKM come straightforwardly from gauge curvatures, that is, KK masses have a gauge origin and in that sense we may call them {\it gauge masses}.


Consider an effective field theory that is characterized by the action
\begin{equation}
\label{A1}
S[{\cal A}^a_M]=\int d^{4+n}x \left(-\frac{1}{4}{\cal F}^a_{MN}(x,\bar x){\cal F}^{MN}_a(x,\bar x)+\sum_{k=1}^\infty\sum_{j_k}\frac{\alpha_{j_k}}{\Lambda^k}{\cal L}_{j_k}({\cal F},{\cal D}{\cal F})\right)\, ,
\end{equation}
defined on the $(4+n)$-dimensional spacetime manifold ${\cal M}^{4+n}={\cal M}^4\times{\cal N}^n$, with ${\cal M}^4$ the usual 4-dimensional spacetime and ${\cal N}^n$ an $n$-dimensional spatial flat manifold. This theory is governed by the Poincar\'e group ${\rm ISO}(1,3+n)$ and the gauge group $SU(N,{\cal M}^{4+n})$.
Here, $d^{4+n}x=d^4x \, d^n{\bar x}$, $x\in {\cal M}^4$, $\bar x \in {\cal N}^n$, and $M=\mu, \bar \mu$, with $\mu=0,1,2,3$ and $\bar \mu=5,\cdots, 4+n$. We use $g_{MN}={\rm diag}(+1,-1,\cdots,-1)$. The Yang-Mills (YM) curvatures ${\cal F}^a_{MN}(x,\bar x)$ are written in terms of the $SU(N,{\cal M}^{4+n})$ connections, ${\cal A}^a_M(x,\bar{x})$, and the dimensionful coupling constant $g_{(4+n)}$. This extra-dimensional description is not renormalizable in the Dyson's sense, so it includes an infinite number of gauge-invariant terms ${\cal L}_{j_k}$, of all possible canonical dimensions $d=4+n+k$ and which are suppressed by inverse powers of an unknown high-energy scale $\Lambda$. The building blocks of the whole theory are YM curvatures and covariant derivatives operating on them. The terms of lowest canonical dimension form a $(4+n)$-dimensional replica of the YM theory in 4 dimensions. Throughout the present paper, we center our discussion in this extra-dimensional YM theory. Experiments are compatible with extra-dimensional theories if the extra dimensions are small enough, which defines a {\it compactification scale}, here denoted by $R^{-1}$. At high energies, far above $R^{-1}$, the action $S[{\cal A}^a_M]$ is governed by the symmetry group ${\rm ISO}(1,3+n)$. We assume that the sizes of the extra dimensions are so large compared with the distance domain explored at such energies that they can be considered infinite. To describe physical phenomena at lower energies, where the compactness of the extra dimensions becomes apparent, we need to {\it hide the symmetry} ${\rm ISO}(1,3+n)$ into ${\rm ISO}(1,3)$, by which we do not mean moving from one theory to another but rather focusing on the same theory from another perspective. Thus ${\rm ISO}(1,3+n)\mapsto {\rm ISO}(1,3)$ occurs through canonical transformations. It is opportune emphasizing that {\it the hiding of the symmetry ${\rm ISO}(1,3+n)$ into ${\rm ISO}(1,3)$ must be compactification-scheme independent}. Hidden symmetries are common to all theories in which a map $G\mapsto H$, from a group $G$ to any non-trivial subgroup $H\subset G$ occurs, as it is the case of the EHM in the SM and, in general, in SSB. In conventional SM extensions, canonical maps that hide symmetries connect two different Lie groups, both defined on the same spacetime manifold. In extra-dimensional theories the map occurs between two different spacetime manifolds, but preserve Lie gauge groups. So, $SU(N,{\cal M}^{4+n})$ and $SU(N,{\cal M}^4)$ coincide as Lie groups, but differ as gauge groups, since their connections are different. Moreover, from the ${\rm ISO}(1,3)$ viewpoint  some connections of the extra-dimensional gauge group appear in tensor representations, which shall be quite relevant for our discussion, below, on mass generation.

To carry out the program of hidding the symmetry, we start by mapping covariant objets of ${\rm SO}(1,3+n)$ into covariant objects of its subgroups ${\rm SO}(1,3)$ and ${\rm SO}(n)$: ${\rm SO}(1,3+n)\mapsto\{ {\rm SO}(1,3),\, {\rm SO}(n)\}$, which means
\begin{eqnarray}
&{\cal A}^a_M(x,\bar x)\mapsto \{{\cal A}^a_\mu(x,\bar x), \, {\cal A}^a_{\bar \mu}(x,\bar x) \}\, ,
\label{M1}
\\
&{\cal F}^a_{MN}(x,\bar x) \mapsto \{ {\cal F}^a_{\mu \nu}(x,\bar x), \, {\cal F}^a_{\mu \bar \nu }(x,\bar x), \, {\cal F}^a_{\bar \mu \bar \nu}(x,\bar x)\} \, ,
\label{M1F}
\end{eqnarray}
with ${\cal A}^a_\mu(x,\bar x)$ and ${\cal A}^a_{\bar \mu}(x,\bar x)$ transforming, respectively, as 1-form (0-form) and 0-form (1-form) under ${\rm SO}(1,3)\,({\rm SO}(n))$. Moreover, ${\cal F}^a_{\mu \nu}(x,\bar x)$, ${\cal F}^a_{\mu \bar \nu }(x,\bar x)$, and ${\cal F}^a_{\bar \mu \bar \nu}(x,\bar x)$ respectively transform as 2-form (0-form), 1-form (1-form), and 0-form (2-form) under ${\rm SO}(1,3)\,({\rm SO}(n))$. At the phase space, the point transformation~(\ref{M1}) is elevated to a canonical transformation. The YM action term, which is the first term of Eq.~(\ref{A1}), becomes
 \begin{equation}
 \label{A2}
 S_{\rm YM}=-\frac{1}{4}\int d^{4+n}x \Big({\cal F}^a_{\mu \nu}{\cal F}^{a\mu \nu}-2{\cal F}^a_{\mu \bar \nu}{\cal F}^{a\mu}\hspace{0.0000001cm}_{\bar \nu}
 +{\cal F}^a_{\bar \mu \bar \nu}{\cal F}^{a\bar \mu \bar \nu}\Big)\, .
 \end{equation}
To move completely from ${\rm ISO}(1,3+n)$ to ${\rm ISO}(1,3)$ we need to get rid of all $\bar x$-coordinate dependence from the theory, which is nontrivial because in the original theory these coordinates are labels that count degrees of freedom. To that aim, we define another canonical map that hides any manifest dynamical role of the ${\rm ISO}(n)$ subgroup at low energies. We assume that some compactification procedure on the ${\cal N}^n$ submanifold has been carried out, and we let $\{ f^{(\underline{m})}(\bar x)\}$ be a complete set of orthogonal functions that are defined on the compact manifold. We develop our discussion by maintaining the set $\{{f^{(\underline{m})}(\bar{x})}\}$ as general as possible, but keep in mind that, as we discuss below, any particular set is associated to boundary conditions that characterize the geometry of the compact manifold. So, any particular choice of this set is, to some extent, equivalent to pick an specific geometry of the compact manifold. In this context, the connections and gauge parameters, $\alpha^a(x,\bar x)$, are expanded as
\begin{equation}
\label{MS}
\varphi_A(x,\bar x)=\sum_{(\underline{m})}f^{(\underline{m})}(\bar x)\varphi^{(\underline{m})}_A(x)\, ,
\end{equation}
where $\varphi_A(x,\bar x)={\cal A}^a_\mu(x,\bar x), \, {\cal A}^a_{\bar \mu}(x,\bar x), \,  \alpha^a(x,\bar x)$, while the 4-dimensional components along the direction $f^{(\underline{m})}(x)$, commonly referred to as {\it KK modes}, are $\varphi_A^{(\underline{m})}(x)= A^{(\underline{m})a}_\mu(x), \, A^{(\underline{m})a}_{\bar \mu}(x), \,  \alpha^{(\underline{m})a}(x)$.
In this expression, we have introduced the {\it KK index}, which we denote by $(\underline{m})=(\underline{m}_1,\underline{m}_2,\ldots,\underline{m}_n)$. Any underlined variable $\underline{m}_j$ represents a discrete index that can be either 0 or a natural number. The symbol $\sum_{(\underline{m})}$, explicitly defined in Refs.~\cite{LMNTedym}, denotes a sum over all possible KK indices $(\underline{m})$.
Since we assumed that the set of basis functions is complete, this map can be reversed as
\begin{equation}
\varphi_A^{(\underline{m})}(x)=\int d^n\bar{x}\,f^{(\underline{m})}(\bar{x})\,\varphi_A(x,\bar{x}),
\end{equation}
which determines the fields that represent the degrees of freedom of the theory, after the canonical transformation.
Thus in the right-hand side of map (\ref{MS}) the degrees of freedom are characterized by the infinite set of fields $\{ A^{(\underline{m})a}_\mu, \, A^{(\underline{m})a}_{\bar \mu}\}$, while the functions $f^{(\underline{m})}(\bar x)$ do not represent degrees of freedom. From Eq.~(\ref{MS}), the fields $A^{(\underline{m})a}_\mu$ and  $A^{(\underline{m})a}_{\bar \mu}$ can be shown to transform, respectively, as a vector and as $n$ scalars of ${\rm SO}(1,3)$. Under map (\ref{MS}), the fundamental Poisson's brackets  $\{\varphi_A(x,\bar x), \, \Pi_B(x',\bar x') \}=\delta_{AB}\,\delta(\mathbf{x}-\mathbf{x'})\, \delta(\bar x-\bar x')\,$ become $\{\varphi^{(\underline{m})}_A(x), \, \pi^{(\underline{n})}_B(x') \}=\delta_{AB}\,\delta^{(\underline{m})(\underline{n})}\delta(\mathbf{x}-\mathbf{x'})$, thus showing that this is a canonical map. This result is compactification-scheme independent; it only depends on the completeness of the set $\{f^{(\underline{m})}(\bar x)\}$.

An assumption that has to be made in order to connect the new physics with the low-energy description is that $\{f^{(\underline{m})}(\bar{x})\}$ includes the constant function, here denoted by $f^{(\underline{0})}$. As we show in a moment, the presence of this function, which is the only one that lacks information about the geometric structure of the compact manifold, is crucial because $f^{(\underline{0})}$ defines the low-energy dynamic variables and the gauge parameters that define the gauge transformations in 4 dimensions. To complete the change of perspective, from $(4+n)$ dimensions to 4 dimensions, the 4-dimensional $SU(N,{\cal M}^4)$-covariant objects must be identified. Remarkably, the $SU(N,{\cal M}^4)$ transformations follow from the inclusion of the constant function $f^{(\underline{0})}$. We start from the $(4+n)$-dimensional infinitesimal transformations $\delta {\cal A}^a_M(x,\bar x)={\cal D}^{ab}_M\alpha^b(x,\bar x)$, and then we implement both canonical transformations to get, with the aid of the orthogonality of $\{ f^{(\underline{m})}(\bar{x}) \}$, the 4-dimensional transformations
\begin{eqnarray}
\delta A^{(\underline{0})a}_\mu={\cal D}^{(\underline{0})ab}_\mu\alpha^{(\underline{0})a}-gf^{abc}\sum_{(\underline{m})}A^{(\underline{m})c}_\mu\alpha^{(\underline{m})b},
\label{gtr1}
\\
\delta A^{(\underline{m})a}_\mu=gf^{abc}A^{(\underline{m})b}_\mu\alpha^{(\underline{0})c}+\sum_{(\underline{r})}{\cal D}_\mu^{(\underline{mr})ab}\alpha^{(\underline{r})b}
\label{gtr2}
\\
\delta A^{(\underline{0})a}_{\bar{\mu}}=gf^{abc}A^{(\underline{0})b}_{\bar{\mu}}\alpha^{(\underline{0})c}-\sum_{(\underline{m})}\Big(gf^{abc}
A^{(\underline{m})c}_{\bar{\mu}}
-\delta^{ab}\int d^n\bar{x}f^{(\underline{0})}\partial_{\bar{\mu}}f^{(\underline{m})}\Big)\alpha^{(\underline{m})b},
\label{gtr3}
\\
\delta A^{(\underline{m})a}_{\bar{\mu}}=gf^{abc}A^{(\underline{m})b}_{\bar{\mu}}\alpha^{(\underline{0})c}-gf^{abc}A^{(\underline{0})c}_{\bar{\mu}}\alpha^{(\underline{m})b}-\sum_{(\underline{r})}{\cal D}^{(\underline{mr})ab}_{\bar{\mu}}\alpha^{(\underline{r})b},
\label{gtr4}
\end{eqnarray}
where $g=f^{(\underline{0})}g_{(4+n)}$ is identified as the dimensionless coupling constant. In these expressions, ${\cal D}^{(\underline{0})ab}_\mu=\delta^{ab}\partial_\mu-gf^{abc}A^{(\underline{0})c}_\mu$, whereas the definitions of the differential operator ${\cal D}^{(\underline{mr})ab}_\mu$ and the object ${\cal D}^{(\underline{mr})ab}_{\bar{\mu}}$can be found in Refs.~\cite{LMNT,LMNTedym,NoTo1}. If we set $\alpha^{(\underline{m})a}=0$ in Eqs.~(\ref{gtr1}) to (\ref{gtr4}), the parameters $\alpha^{(\underline{0})a}$, along the direction of $f^{(\underline{0})}$, are identified as the gauge parameters that define the $SU(N,{\cal M}^4)$ transformations under which $A^{(\underline{0})a}_\mu$ transforms as a gauge field, and $A^{(\underline{m})a}_\mu$, $A^{(\underline{0})a}_{\bar{\mu}}$, $A^{(\underline{m})a}_{\bar{\mu}}$ transform as matter fields. We call these transformations the {\it standard gauge transformations}. Similarly, the parameters $\alpha^{(\underline{m})a}(x)$ define a set of {\it non-standard gauge transformations}, which is inherent to the canonical transformation implemented to hide gauge symmetry, but which manifestly shows that there is a larger gauge symmetry underlying the KK description. Within the BRST formalism~\cite{BRST}, these parameters are
central to KK quantization~\cite{NoTo1,GNT}.

So far there is, besides the $SU(N,{\cal M}^4)$ {\it standard} gauge field $A^{(\underline{0})a}_\mu$, a set of $n$ low-energy scalars $A^{(\underline{0})a}_{\bar{\mu}}$, which we wish to eliminate in order to ensure that the low-energy theory is no other than the YM theory in four dimensions. Inspired by the evenness of $f^{(\underline{0})}$ under $\bar{x}\to-\bar{x}$, we assume that the compactification scheme is such that the whole set $\{ f^{(\underline{m})}(\bar{x}) \}$  can be split into disjoint subsets of even functions,  $\{ f^{(\underline{0})},f^{(\underline{m})}_E(\bar{x}) \}$, and odd functions, $\{f^{(\underline{m})}_O(\bar{x}) \}$. We postulate that any field with standard counterpart is even, while fields without standard counterpart are odd. Then we assume that the vectors ${\cal A}^a_{\mu}$ are even, while the scalars ${\cal A}^a_{\bar{\mu}}$ are odd, and we assume that the gauge parameters $\alpha^a$ are even. The canonical map~(\ref{MS}) then reads
\begin{eqnarray}
{\cal A}^a_\mu(x,\bar{x})&=&f^{(\underline{0})}A^{(\underline{0})a}_\mu(x)+\sum_{(\underline{m})}f^{(\underline{m})}_E(\bar{x})A^{(\underline{m})a}_\mu(x),
\\
{\cal A}^a_{\bar{\mu}}(x,\bar{x})&=&\sum_{(\underline{m})}f^{(\underline{m})}_O(\bar{x})A^{(\underline{m})a}_{\bar{\mu}}(x),
\\
\alpha^a(x,\bar{x})&=&f^{(\underline{0})}\alpha^{(\underline{0})a}(x)+\sum_{(\underline{m})}f^{(\underline{m})}_E(\bar{x})\alpha^{(\underline{m})a}(x),
\end{eqnarray}
where the KK sums run over all KK indices, except for $(\underline{0})$.
Clearly, the parity assignments leave only scalars $A^{(\underline{m})a}_{\bar \mu}$ with $(\underline{m})\ne(\underline{0})$. The fields $A^{(\underline{m})a}_{ \mu}$ and $A^{(\underline{m})a}_{\bar \mu}$ are known as the {\it KK excitations of} $A^{(\underline{0})a}_{ \mu}$.

Now we discuss the mass-generating KKM. The $SU(N,{\cal M}^4)$ symmetry prevents the zero mode $A^{(\underline{0})}_\mu$ from acquiring mass, but $A^{(\underline{m})a}_\mu$ and the scalars $A^{(\underline{m})a}_{\bar \mu}$ are tensorial representations of $SU(N,{\cal M}^4)$, so their masses are allowed by this symmetry. Such masses must be proportional to the compactification scale $R^{-1}$ in order for any physical effect induced by these fields to decouple when $R^{-1}\gg v$~\cite{AppCa}.
The $(4+n)$-dimensional curvature components given in Eq.~(\ref{M1F}) inherit definite parities from the $(4+n)$-dimensional fields ${\cal A}^a_\mu$ and ${\cal A}^a_{\bar{\mu}}$, so they are mapped as
\begin{eqnarray}
{\cal F}^a_{\mu\nu}(x,\bar{x})&=&f^{(\underline{0})}{\cal F}^{(\underline{0})a}_{\mu\nu}(x)+\sum_{(\underline{m})}f^{(\underline{m})}_E(\bar{x}){\cal F}^{(\underline{m})a}_{\mu\nu}(x),
\\
{\cal F}^a_{\mu\bar{\nu}}(x,\bar{x})&=&\sum_{(\underline{m})}f^{(\underline{m})}_O(\bar{x}){\cal F}^{(\underline{m})a}_{\mu\bar{\nu}}(x),
\\
{\cal F}^a_{\bar{\mu}\bar{\nu}}(x,\bar{x})&=&f^{(\underline{0})}{\cal F}^{(\underline{0})}_{\bar{\mu}\bar{\nu}}(x)+\sum_{(\underline{m})}f^{(\underline{m})}_E(\bar{x}){\cal F}^{(\underline{m})a}_{\bar{\mu}\bar{\nu}}(x).
\end{eqnarray}
Then, integrating out the $\bar{x}$ coordinates in the YM action (\ref{A2}) yields the effective Lagrangian $
{\cal L}^{\rm YM}_{\textrm{eff}}={\cal L}^{\rm YM}_{\textrm{v-v}}+{\cal L}^{\rm YM}_{\textrm{v-s}}+{\cal L}^{\rm YM}_{\textrm{s-s}}$, with
\begin{eqnarray}
{\cal L}^{\rm YM}_{\textrm{v-v}}&=&-\frac{1}{4}{\cal F}^{(\underline{0})a}_{\mu\nu}{\cal F}^{(\underline{0})a\mu\nu}-\frac{1}{4}\sum_{(\underline{m})}{\cal F}^{(\underline{m})a}_{\mu\nu}{\cal F}^{(\underline{m})a\mu\nu},
\label{Lvv}
\\
\label{KS}
{\cal L}^{\rm YM}_{\textrm{v-s}}&=&\frac{1}{2}\sum_{(\underline{m})}{\cal F}^{(\underline{m})a}_{\mu \bar \nu}(x){\cal F}^{(\underline{m})a\mu}_{ \bar \nu}(x)\, , \\
\label{SP}
{\cal L}^{\rm YM}_{\textrm{s-s}}&=&-\frac{1}{4}{\cal F}^{(\underline{0})a}_{\bar \mu \bar \nu}{\cal F}^{(\underline{0})a \bar \mu \bar \nu}-\frac{1}{4}\sum_{(\underline{m})}{\cal F}^{(\underline{m})a}_{\bar \mu \bar \nu}{\cal F}^{(\underline{m})a \bar \mu \bar \nu}\, .
\end{eqnarray}
The KK modes of the extra-dimensional curvatures are given by
\begin{eqnarray}
{\cal F}^{(\underline{0})a}_{\mu\nu}&=&F^{(\underline{0})a}_{\mu\nu}+gf^{abc}\sum_{(\underline{m})}A^{(\underline{m})b}_\mu A^{(\underline{m})c}_\nu,
\\
{\cal F}^{(\underline{m})a}_{\mu\nu}&=&{\cal D}^{(\underline{0})ab}_\mu A^{(\underline{m})b}_\nu-{\cal D}^{(\underline{0})ab}_\nu A^{(\underline{m})b}_\mu+gf^{abc}\sum_{(\underline{rs})}\Delta_{(\underline{mrs})}A^{(\underline{r})b}_\mu A^{(\underline{s})c}_\nu,
\\
{\cal F}^{(\underline{m})a}_{\mu\bar{\nu}}&=&{\cal D}^{(\underline{0})ab}_\mu A^{(\underline{m})b}_{\bar{\nu}}-\sum_{(\underline{r})}p_{\bar{\nu}}^{(\underline{mr})} A^{(\underline{r})a}_\mu+gf^{abc}\sum_{(\underline{rs})}\Delta'_{(\underline{mrs})}A^{(\underline{s})b}_\mu A^{(\underline{r})c}_{\bar{\nu}},
\label{vscrv}
\\
{\cal F}^{(\underline{0})a}_{\bar{\mu}\bar{\nu}}&=&gf^{abc}\sum_{(\underline{m})}A^{(\underline{m})b}_{\bar{\mu}}A^{(\underline{m})c}_{\bar{\nu}},
\\
{\cal F}^{(\underline{m})a}_{\bar{\mu}\bar{\nu}}&=&\sum_{(\underline{r})}(p^{(\underline{mr})}_{\bar{\mu}}\delta_{\bar{\nu}\bar{\alpha}}-p^{(\underline{mr})}_{\bar{\nu}}\delta_{\bar{\mu}\bar{\alpha}})A^{(\underline{r})a}_{\bar{\alpha}}+gf^{abc}\sum_{(\underline{rs})}\Delta'_{(\underline{rsm})}A^{(\underline{r})b}_{\bar{\mu}}A^{(\underline{s})c}_{\bar{\nu}}.
\label{sscrv}
\end{eqnarray}
Here, $F^{(\underline{0})}_{\mu\nu}$ is the $SU(N,{\cal M}^4)$ curvature (not a KK mode!), in terms of the connections $A^{(\underline{0})a}_\mu$, meaning that ${\cal L}^{\rm YM}_{\textrm{v-v}}$, Eq.~(\ref{Lvv}), contains the 4-dimensional YM Lagrangian, which is the low-energy theory. Besides this, ${\cal L}^{\rm YM}_{\textrm{v-v}}$ comprises the kinetic terms for the vector KK modes $A^{(\underline{m})a}_\mu$ and couplings of these matter fields with the connections. In this sense, ${\cal L}^{\rm YM}_{\textrm{v-v}}$ evokes the YM sector of the electroweak SM, after SSB, where the $U_e(1)$ field and the massive $W$ gauge boson are, respectively, analogues of $A^{(\underline{0})a}_\mu$ and $A^{(\underline{m})a}_\mu$. The factors $\Delta_{(\underline{rsm})}$ and $\Delta'_{(\underline{rsm})}$, which are sums of terms constituted by products of Kronecker deltas, are properly defined in Ref.~\cite{LMNTedym}. Furthermore, we have the object
\begin{equation}
p^{(\underline{mr})}_{\bar{\nu}}=\int d^n\bar{x}\,f_O^{(\underline{m})}(\bar{x})\partial_{\bar{\nu}}f_E^{(\underline{r})}(\bar{x}),
\label{edpf}
\end{equation}
which appears in terms that are linear in the fields $A^{(\underline{m})a}_\mu$, in Eq.~(\ref{vscrv}), and $A^{(\underline{m})a}_{\bar{\mu}}$, in Eq.~(\ref{sscrv}). Since KK curvatures lie quadratically in the terms that constitute the KK Lagrangian ${\cal L}^{\rm YM}_{\rm eff}$, it turns out that $p^{(\underline{mr})}_{\bar{\nu}}$, which does not involve degrees of freedom, essentially determines the mass spectrum, once the set $\{ f^{(\underline{m})}(\bar{x}) \}$ has been chosen. Nevertheless, note that we do not need to define a particular geometry of extra dimensions to generate KK mass terms. It is worth emphasizing that no mass terms for the gauge field $A^{(\underline{0})a}_\mu$ are generated, so it consistently remains massless. Since the mass terms for the KK vectors $A^{(\underline{m})a}_\mu$ emerge from Eq.~(\ref{KS}) and the masses of KK scalars $A^{(\underline{m})a}_{\bar{\mu}}$ come from Eq.~(\ref{SP}), we note that ${\cal L}^{\rm YM}_{\textrm{v-s}}$ and ${\cal L}^{\rm YM}_{\textrm{s-s}}$ are analogues of the kinetic term and the scalar potential of the Higgs sector in the SM, respectively. Such a resemblance is further established by inspecting the structure of the couplings in these lagrangian terms\footnote{A detailed discussion, including explicit expressions, will appear elsewhere~\cite{GLMNNT}.}. In the SM and in a variety of its extensions, masses are generated by SSB, but to do so the addition of more degrees of freedom to the theory is required. For instance, in the SM a Higgs doublet, defining a whole scalar sector, is introduced and, by the use of an appropriate scalar potential, gauge symmetry is broken, thus defining the SM masses. By contrast, in the KKM masses have a gauge origin: from the sole extra-dimensional curvatures, and without introducing extra degrees of freedom, compactification produces mass terms.

The structure of Eq.~(\ref{edpf}) suggests that the elements of $\{f^{(\underline{m})}(\bar{x})\}$ might be eigenfunctions of the differential operator $(-i\,\partial_{\bar{\nu}})^2=-\bar{\nabla}^2$, whose representation in the ket space is the Casimir invariant $P_{\bar{\nu}}P_{\bar{\nu}}=\bar{P}^2$, associated to the inhomogeneous translations group ${\rm ISO}(n)$, with eigenkets $| \bar{p} \rangle$: $\bar{P}^2| \bar{p} \rangle=\bar{p}^2| \bar{p} \rangle$, where $\bar{p}$ is an extra-dimensional momentum eigenvalue. The extra-dimensional momentum operator $P_{\bar{\nu}}$ is associated to an observable, so it is hermitian. Thus its eigenvalues $p_{\bar{\nu}}$ are real numbers, which means that the eigenvalues $\bar{p}^2=p_{\bar{\nu}}p_{\bar{\nu}}$, of our Casimir invariant, are positive quantities; this is a necessary requirement for $p_{\bar{\nu}}^{(\underline{mr})}$, Eq.~(\ref{edpf}), to consistently define the KK masses. Our criterion to choose a set of eigenfunctions $\{ f^{(\underline{0})},f^{(\underline{m})}_E(\bar{x}),f^{(\underline{m})}_O(\bar{x}) \}$ consists in the selection of an extra-dimensional observable, in this case the aforementioned Casimir invariant. Thus we identify $f^{(\underline{m})}(\bar{x})=\langle \bar{x} | \bar{p}^{(\underline{m})} \rangle$ as wave eigenfunctions satisfying the Laplace equation $\bar{\nabla}^2f^{(\underline{m})}(\bar{x})=-\bar{p}^2f^{(\underline{m})}(\bar{x})$, for momenta defined in the compact manifold ${\cal N}^n$, with plane-wave solutions. At this point, we implement a compactification scheme, for which we assume that each coordinate axis $\bar x^j$ is coiled in a circle $S^1$ of radius $R_j$. Since we need to define parity on the manifold, we introduce the orbifold $S^1/Z_2$. Then, we assume that ${\cal N}^n$ is made of $n$ copies of the orbifold $S^1/Z_2$, that is, ${\cal N}^n=(S^1/Z_2)\otimes\cdots\otimes(S^1/Z_2)$, with the orbifolds having radii $R_1,R_2,\ldots, R_n$. Dirichlet and Neumann boundary conditions respectively yield $f^{(\underline{m})}_E(\bar{x})=\cos\left(\bar{x}\cdot\bar{p}^{(m)}\right)$ and $f^{(\underline{m})}_O(\bar{x})=\sin\left(\bar{x}\cdot\bar{p}^{(m)}\right)$, where $p^{(\underline{m})}=\left( \underline{m}_1/R_1,\ldots,\underline{m}_n/R_n \right)^{\rm T}$. With this in mind, we define the quantity $m^2_{(\underline{m})}\equiv\bar{p}^{(\underline{m})}_{\bar{\nu}}\bar{p}^{(\underline{m})}_{\bar{\nu}}=\sum_{k=1}^n(\underline{m}_k/R_k)^2$, which characterizes the eigenvalue spectrum of $\bar{P}^2$. For simplicity, we take all radii equal, $R_1=R_2=\ldots=R_n\equiv R$, in which case $m_{(\underline{m})}^2=R^{-2}\underline{m}^2$.

Our definition of $\{ f^{(\underline{m})}(\bar x) \}$  leaded us to a concrete expression of $p^{(\underline{mr})}_{\bar{\nu}}$, Eq.~(\ref{edpf}), and so it completes the definitions of the KK curvatures ${\cal F}^{(\underline{m})a}_{\mu\bar{\nu}}$ and ${\cal F}^{(\underline{m})a}_{\bar{\mu}\bar{\nu}}$, thus determining the mass spectrum that originates in the KKM. The masses for the gauge fields $A^{(\underline{m})a}_\mu$, nested in the lagrangian ${\cal L}^{\rm YM}_{\textrm{v-s}}$, are straightforwardly found to be $m_{(\underline{m})}$, but the determination of the KK scalar masses deserves further commenting. Bilinear scalar terms, conceived within the second term of ${\cal L}^{\rm YM}_{\textrm{s-s}}$, show up as mixings of scalars $A^{(\underline{m})a}_{\bar{\mu}}$. These mixings occur through $n\times n$ symmetric mass matrices (one per each KK index $(\underline{m})$) $\mathfrak{M}_{\bar{\mu}\bar{\nu}}^{(\underline{m})}=m^2_{(\underline{m})}\delta_{\bar{\mu}\bar{\nu}}-p^{(\underline{m})}_{\bar{\mu}}p^{(\underline{m})}_{\bar{\nu}}$. Diagonalization of $\mathfrak{M}^{(\underline{m})}_{\bar{\mu}\bar{\nu}}$, by which the KK scalars are lead to the mass-eigenstates basis, yields $\mathfrak{M}^{(\underline{m})}_{\bar{\mu}\bar{\nu}}\to{\rm diag}(m^2_{(\underline{m})},m^2_{(\underline{m})},\ldots, m^2_{(\underline{m})},0)$, which reveals a quite relevant aspect of the set of KK scalar fields: while in the mass-eigenstates basis $(n-1)$ KK scalars $A^{(\underline{m})a}_{\bar{n}}$ ($\bar{n}=1,2,\ldots,n-1$) have mass $m_{(\underline{m})}$, one of them, $A^{(\underline{m})a}_G$, turns out to be massless. SM gauge-boson masses, originated in the EHM, are given as the product of a dimensionless constant and a dimensionful energy scale, in this case the vacuum expectation value $v$. Kaluza-Klein masses, $m_{(\underline{m})}=R^{-1}(\underline{m}^2)^{1/2}$, generated by the KKM, share this feature, as they are given as the product of dimensionless constants $(\underline{m}^2)^{1/2}$ and the compactification scale $R^{-1}$, with mass units. The massless scalar $A^{(\underline{m})a}_G$ behaves as a pseudo-Goldstone boson, in the sense that it is completely removed from the theory by fixing the gauge as $\alpha^{(\underline{m})a}=A^{(\underline{m})a}_G/m_{(\underline{m})}$, with respect to the non-standard gauge transformations, which resembles the case of the unitary gauge in the Higgs sector of the SM. All our discussion, and in particular that on the KKM, has been performed around the YM theory. In the richer context of the SM (4+n)-dimensional extension, SSB happens in addition to the KKM, and fields with KK masses receive a further contribution from the EHM, thus defining masses given by $m^2_{\varphi^{(\underline{m})}}=m^2_{\varphi^{(\underline{0})}}+m^2_{(\underline{m})}$ for the KK excitations $\varphi^{(\underline{m})}$, of the low-energy dynamic variables $\varphi^{(\underline{0})}$, whose mass is $m_{\varphi^{(\underline{0})}}$.

In a theory that involves a Higgs sector, with SSB implemented by a constant vector $\Phi_{0}$ that defines a point in a minimal-energy hypersurface, a gauge group $G$, of dimension $d_G$, is broken down to some subgroup $H$, of dimension $d_H$, resulting in $d_G-d_H$ broken generators that are in one-to-one correspondence with $d_G-d_H$ gauge fields acquiring masses and $d_G-d_H$ gauge parameters defining nonstandard gauge transformations. Collaterally, $d_G-d_H$ pseudo-Goldstone bosons arise. 
Any massive gauge boson points towards the direction of a broken generator of $G$, whereas the remaining $d_H$ gauge fields of $H$ are aligned with the directions of $d_H$ unbroken generators of $G$, which define the Lie algebra of $H$. In KK theories, the Casimir invariant ${\bar P}^2$ defines the discrete basis $\{ | p^{(\underline{m})} \rangle \}$. An analogy between this eigenket basis and the aforementioned basis of gauge Lie-group generators can be established. Each eigenket $|p^{(\underline{m})}\neq 0\rangle $ is the analogue of a broken gauge-group generator of $G$, although the set of momentum eigenkets involves an infinite number of directions, while the set of generators of a gauge Lie group is finite. In fact, associated with each $|p^{(\underline{m})}\neq 0\rangle $ there is a massive gauge field $A^{(\underline{m})a}_\mu$, a gauge parameter $\alpha^{(\underline{m})a}$ defining nonstandard gauge transformations, and a pseudo-Goldstone boson $A^{(\underline{m})a}_G$. In this analogy, the ``base state''  $|0\rangle $ plays the role of an unbroken gauge-group generator in the sense that a gauge field pointing in this direction remains massless.

To construct the eigenket basis $\{|p^{(\underline{m})}\rangle\}$, we used the Casimir invariant associated with the translations subgroup of ${\rm ISO}(n)$, which was appropriate for practical purposes. In a broader context, the $(n+1)/2$ Casimir invariants of this group~\cite{CDN} can be used to construct a more general basis, with eigenkets $|p^{(\underline{m})}, \sigma \rangle$, where $\sigma$ labels, collectively, other eigenvalues. It is then clear that the ${\rm ISO}(n)$ generators $P_{\bar \mu}$, $J_{\bar \mu \bar \nu}$ and the generators $J_{\mu \bar \nu}$ (no group), which are ${\rm ISO}(1,3+n)$ generators that do not belong to ${\rm ISO}(1,3)$, are broken in the sense that $(P_{\bar \mu}, J_{\bar \mu \bar \nu}, J_{\mu \bar \nu})|p^{(\underline{m})}, \sigma \rangle \neq 0$, whereas the ${\rm ISO}(1,3)$ generators are unbroken: $(P_\mu, J_{\mu \nu})|p^{(\underline{m})}, \sigma \rangle = 0$. Independently of broken generators, in our basis there is a one-to-one correspondence between eigenkets $| p^{(\underline{m})} \rangle$ and triplets $(A^{(\underline{m})a}_\mu,A^{(\underline{m})a}_G, \alpha^{(\underline{m})a})$, in the sense of Fourier multi-indices. The KK massive gauge fields lie along the base kets $| p^{(\underline{m})} \rangle$. In both SSB and the KKM, the physical content is a matter of scales: at energies of order $R^{-1}$, we use the ${\rm ISO}(1,3)$ description, but ${\rm ISO(1,4+n)}$ is used at energies far above of $R^{-1}$, which corresponds to distances so small that compact dimensions seem infinite.


This letter has been devoted to discuss the concepts of hidden symmetry and the mass-generating Kaluza-Klein mechanism, both of which incarnate essential elements of Kaluza-Klein effective theories, whose origins are field formulations in spacetimes with extra dimensions. We argued that symmetries in extra dimensions are hidden through canonical transformations ${\rm ISO}(1,3+n)\mapsto\{{\rm ISO}(1,3),{\rm ISO}(n)\}\mapsto{\rm ISO}(1,3)$ that change the extra-dimensional perspective by a description in four dimensions. The dynamic variables of the resulting effective theory are representations of ${\rm SO}(1,3)$, which include an $SU(N,{\cal M}^4)$ gauge field, Kaluza-Klein vector fields, a set of Kaluza-Klein scalars and pseudo-Goldstone bosons. While the hiding of ${\rm ISO}(1,3+n)$ into ${\rm ISO}(1,3)$ and the corresponding definition of 4-dimensional variables, by the second canonical map, is independent of the geometry of the compact extra dimensions, the presence of a constant function in such a change of variables is mandatory in order to establish which are the dynamic variables and gauge parameters that characterize the low-energy theory. From the 4-dimensional viewpoint, some extra-dimensional gauge degrees of freedom behave as matter fields, which, unprotected by gauge symmetry in 4 dimensions, become massive. Kaluza-Klein mass terms naturally emerge from the very gauge structure of the theory in extra-dimensions, without the need of introducing extra degrees of freedom. The resulting mass spectrum arises no matter what the geometry of the compact extra dimensions is. Nevertheless, we showed that the extra-dimensional squared-momentum Casimir invariant, which is an observable associated to the inhomogeneous group of translations in the extra dimensions, provides a set of wave eigenfunctions that, together with the boundary conditions that characterize the geometry of a given compact manifold, sets a Kaluza-Klein mass spectrum that is consistent with the decoupling theorem.

\section*{Acknowledgements}
\noindent
We acknowledge financial support from CONACYT and SNI (M\'exico).

\end{document}